\documentclass[preprints,article,accept,pdftex,moreauthors]{Definitions/mdpi}

\firstpage{1}
\pubvolume{1}
\issuenum{1}
\articlenumber{0}
\pubyear{2025}
\copyrightyear{2025}
\datereceived{ }
\daterevised{ }
\dateaccepted{ }
\datepublished{ }
\hreflink{https://doi.org/}

\usepackage{bm,color,xcolor,amsfonts,amsmath,amssymb,epsfig,esint}

\newcommand{\be}{\begin{equation}}
\newcommand{\ee}{\end{equation}}
\newcommand{\bea}{\begin{eqnarray}}
\newcommand{\eea}{\end{eqnarray}}
\newcommand{\beas}{\begin{eqnarray*}}
\newcommand{\eeas}{\end{eqnarray*}}

\newcommand{\vep}{{\bm p}}

\newcommand{\veq}{{\bm q}}

\newcommand{\ver}{{\bm r}}

\newcommand{\veep}{{\bm \epsilon}}

\newcommand{\X}{X(3872)}

\newcommand{\mS}{M}

\newcommand{\dszero}{D^*_{s0}}
\newcommand{\dsone}{D_{s1}}
\newcommand{\kappaon}{\kappa_{\rm loop}(q^2=m_{\dszero}^2)}
\newcommand{\kappaoff}{\kappa_{\rm loop}(q^2)}

\newcommand{\bonn}{Helmholtz-Institut f\"ur Strahlen- und Kernphysik, Universit\"at Bonn, D-53115 Bonn, Germany}

\newcommand{\fzj}{Institute for Advanced Simulation, Forschungszentrum J\"ulich, D-52425 J\"ulich, Germany}

\newcommand{\itp}{Institute of Theoretical Physics, Chinese Academy of Sciences, Beijing 100190, China}

\newcommand{\ucas}{School of Physical Sciences, University of Chinese Academy of Sciences, Beijing 100049, China}

\graphicspath{{Figs.dir/}}

\Title{Radiative decays of hadronic molecules: \\
From confusion to inspiration}

\TitleCitation{Title}

\Author{Feng-Kun~Guo $^{1,2,}$*\orcidA{}, Christoph~Hanhart $^{3}$\orcidB{} and Alexey Nefediev $^{4}$\orcidC{}}

\AuthorNames{Feng-Kun~Guo, Christoph~Hanhart and Alexey Nefediev}

\address{%
$^{1}$ \quad \itp\\
$^{2}$ \quad \ucas\\
$^{3}$ \quad \fzj\\
$^{4}$ \quad \bonn}

\corres{Correspondence: fkguo@itp.ac.cn}

\abstract{Radiative decays of hadronic states provide an essential source of information that can facilitate deciphering their nature and properties. However, a lot of confusion concerning radiative decays of hadronic molecules and their interpretation can be found in the literature.
In this paper, we briefly review several types of such decays and pinpoint similarities and essential differences between them. In particular, we emphasise the crucial role played by the hierarchy of the scales relevant to the studied system and the resulting necessity of employing an approach that considers them appropriately. We illustrate the situation with several instructive examples.}


\begin{document}



\section{Introduction}

The emission of photons is governed by Quantum Electrodynamics (QED), which is the simplest and most studied field theory within the Standard Model
(SM). For this reason, radiative decays of hadronic states are considered a promising tool to investigate the hadrons in the initial
and final states of the reaction and shed light on their nature and properties. In particular, such decays can provide access to the
inner structure of these hadrons. However, the information provided by radiative decays requires a care  in interpretation to avoid false
conclusions. In this paper, we review several examples of radiative decays of hadronic molecules and pinpoint
various peculiarities related to them. In particular, we discuss the hierarchy of internal scales that can be different for different molecules and address
the impact of the compact component of the wave function of the molecular state on its radiative decays.

The paper is organized as follows: In Sec.~\ref{sec:pos}, we start from a textbook example of the two-photon decay of parapositronium and argue that the famous ``universal'' formula for this decay width expressed in terms of the wave function at the origin, $\psi(0)$, does not apply to hadronic molecules --- states generated by the non-perturbative, strong interactions of hadron pairs, see Ref.~\cite{Guo:2017jvc} for a review. We introduce and discuss an alternative expression for the width derived in the limit of zero-ranged interactions, consistent with the hierarchy of scales relevant for such hadronic molecules and demonstrate that the derived formula works
very well for two-photon decays of the scalar mesons $a_0/f_0(980)$ treated as $K\bar{K}$ molecules. Then, in Sec.~\ref{sec:phi}, we address the radiative decays of $\phi(1020)$ that also involve the above scalars in the final state. We argue how a proper
treatment of the diagrams and the molecular vertex consistent with quantum mechanics and gauge invariance allows one to arrive at a reliable finite prediction for the decay width. In Sec.~\ref{sec:Ds1}, we proceed to the radiative decays of the $\dsone(2460)$ meson treated as a $D^*K$ molecule. In this case, although the loop amplitudes do not diverge, it follows from the power counting that there is no enhancement of the loop diagrams compared with the contact term, so the corresponding contact amplitude needs to be included. Thus, the radiative decays of the $\dsone(2460)$ meson provide an example of the situation when an additional experimental input is necessary to fix the unknown short-range parameter and this way to arrive at a complete theory capable of providing definite predictions for observables. A ratio of the branching fractions of two radiative decays of the $\dsone(2460)$, if measured experimentally, is argued to be able to provide such an input. Finally, in Sec.~\ref{sec:X}, we turn to the radiative decays of the $X(3872)$, also known as $\chi_{c1}(3872)$, as yet another prominent candidate to a hadronic molecule, and argue that the divergence of the loop amplitudes contributing the decays $X\to\gamma\psi$ ($\psi=J/\psi$, $\psi(2S)$) calls for the inclusion of the leading-order short-range contact term that is further found to strongly depend on the renormalisation scale. It is concluded, therefore, that the radiative decays of the $X(3872)$ are mostly sensitive to the short-range component of its wave function and as such are not decisive in probing the molecular nature of the $X$. We conclude in Sec.~\ref{sec:concl}.

\section{Positronium versus a point-like molecule}
\label{sec:pos}

We start from two-photon decays of parapositronium as the most prominent example of the radiative decay of a compound state. Parapositronium is a bound state of the electron-positron pair with the quantum numbers $^1S_0$.
In what follows we also refer to it as ${\rm Ps}(^1S_0)$ or simply ``positronium''. Given its positive $C$-parity, parapositronium decays into an even number of photons, the two-photon decay being the dominant one.
Two features inherent in the positronium make it a distinct object for studies. On the one hand, it is undoubtedly a composite system represented by a generic bound state. On the other hand, the interaction responsible for its binding is very well understood --- the bound state is formed by the attractive Coulomb potential and can be readily found as solution of the corresponding Schr{\"o}dinger equation with the ground state wave function and binding momentum being\footnote{The meaning of the subscript $A$ will be explained below.}
\be
\psi_A(r)=\sqrt{\frac{\gamma^3}{\pi}}\,e^{-\gamma r},\quad
\gamma=\sqrt{2\mu_e E_B} \ ,
\label{psir}
\ee
respectively, where
\be
\gamma=\alpha \mu_e
\quad \mbox{and} \quad \mu_e = \frac{m_e^2}{m_e+m_e}=\frac{1}{2}m_e \ ,
\label{kappapos}
\ee
with $m_e$ the electron mass and $\alpha=e^2/(4\pi)\approx1/137$ the electromagnetic fine structure constant. Hereinafter we use the symbol $\mu$
for the relevant reduced mass. Then, according to textbooks in quantum mechanics, the width of a two-photon decay of the positronium can be evaluated as
\be
\Gamma[{\rm Ps}(^1S_0)\to\gamma\gamma]=v\rho\sigma,
\label{widthPs}
\ee
where $v$ is the velocity of the electron (positron) in the positronium rest frame, $\rho=|\psi_A(0)|^2$ is the probability for the electron and positron to meet each other at the origin (annihilation point), and
\be
\sigma=\frac{4\pi\alpha^2}{m_e^2 v}
\label{sigmaA}
\ee
is the total cross section of the two-photon annihilation of a slow $e^+e^-$ pair in the $^1S_0$ state. Combining Eqs.~\eqref{psir}-\eqref{sigmaA}, it is easy to arrive at the famous formula,
\be
\Gamma[{\rm Ps}(^1S_0)]=\frac{4\pi\alpha^2}{m_e^2}
|\psi_A(0)|^2=\frac12\alpha^5 m_e \ .
\label{widthA}
\ee
The corrections to Eq.~\eqref{widthA} are suppressed by higher powers of the fine structure constant $\alpha$.

\begin{figure}[t!]
\centerline{\epsfig{file=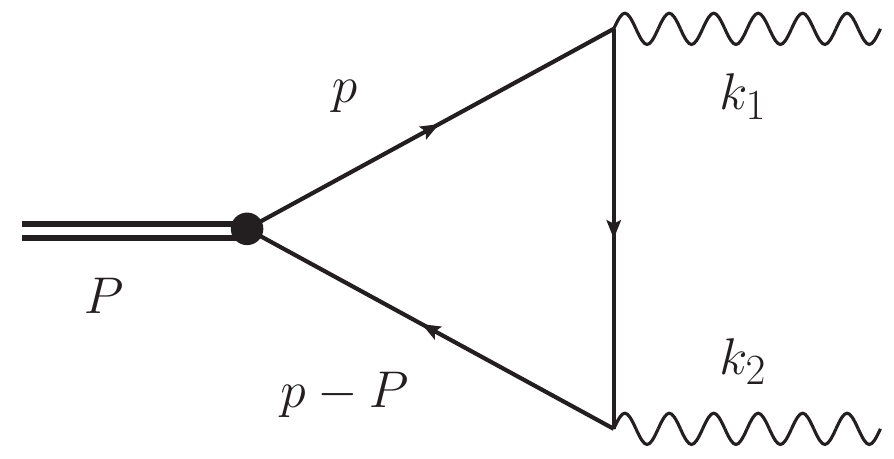,width=0.4\textwidth}\hspace*{0.05\textwidth}
\epsfig{file=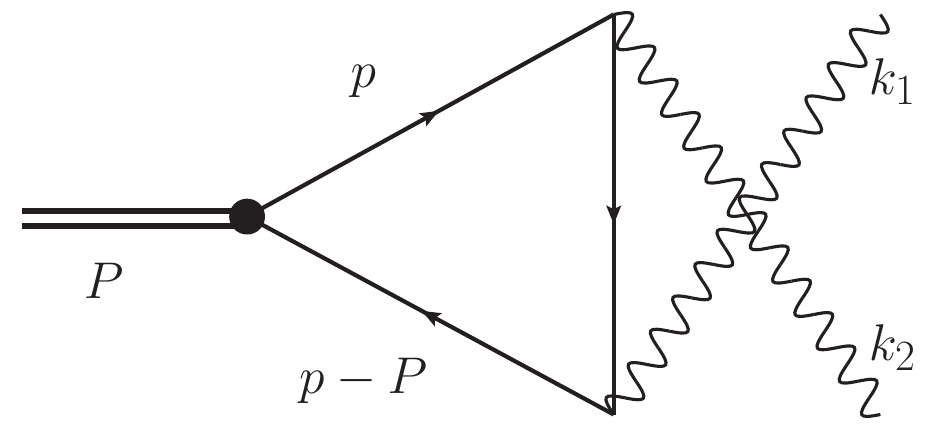, width=0.4\textwidth}}
\caption{The two diagrams contributing to the two-photon decay of parapositronium. The black filled circle indicates the positronium disintegration vertex that can be obtained as solution of the Bethe-Salpeter bound-state equation and contains the positronium wave function as the central (but not only!) ingredient --- see, for example, the derivation of Eq.~\eqref{vert} in Sec.~\ref{sec:phi} below. The other two vertices in the diagrams are the standard photon emission vertices controlled by QED.}\label{fig:posdec}
\end{figure}

The simplicity and the seemingly universal form of Eq.~\eqref{widthA} made it an extremely popular and \emph{de facto} default approach to two-photon decays of compound objects. For example, the two-photon decay of kaonium (the bound state of a kaon-antikaon pair bound by the strong force --- in other words, a hadronic molecule of $K\bar K$) was studied in the literature employing the above technique with results ranging from 0.6~keV \cite{Barnes:1985cy} to 6~keV \cite{Krewald:2003ab}. The order-of-magnitude discrepancy between these predictions should not come as a surprise given that the formula for the decay width relies on the value of the wave function at the origin that is extremely sensitive to the details of the interaction. Since the strong interaction potential of kaons is not known at short distances, any calculation of this kind relies on a particular model, and the corresponding hardly quantifiable model uncertainty automatically propagates to the final result. However,
a blind application of the positronium-like formula \eqref{widthA} to a generic two-photon decay of a molecule is potentially dangerous, since it relies on a particular hierarchy of scales~\cite{Hanhart:2007wa}. Indeed, the positronium two-photon decay amplitude is encoded in the two diagrams depicted in Fig.~\ref{fig:posdec}. Then taking the integral in the loop energy $p_0$, and retaining only the leading nonrelativistic contribution, one readily arrives at
\be
{\cal M}[{\rm Ps}(^1S_0)\to \gamma\gamma]=\frac{1}{\sqrt{m_e}} \int\frac{d^3\vep}{(2\pi)^3}\psi_A(\vep){\cal M}[e^+e^-\to\gamma\gamma],
\label{Tfin}
\ee
where ${\cal M}[e^+e^-\to\gamma\gamma]$ is the amplitude of an unbound electron-positron pair annihilation into two photons. Equation~(\ref{Tfin})
is still general. In the case of positronium
 it is easy to see that the two objects under the momentum integral in the amplitude \eqref{Tfin} possess very different typical momentum scales. While the wave function effectively cuts the integral  at $p\equiv |\vep|\sim \alpha m_e$, the momentum scale controlling the transition to the two photons is much larger, of the order of $m_e$, such that ${\cal M}[e^+e^-\to\gamma\gamma]$ stays basically constant in the momentum range, where
 the integral has already converged.
 As a result, the amplitude ${\cal M}[e^+e^-\to\gamma\gamma]$ can be pulled out from the integral, so the positronium coordinate-space wave function taken at the origin,
\be
\psi_A(0)=\int\frac{d^3\vep}{(2\pi)^3}\psi_A(\vep),
\ee
arises quite naturally. Thus we conclude that the positronium-like formula \eqref{widthA} relies on a particular hierarchy of scales,
\be
r_\Gamma\gg r_a,
\label{caseA}
\ee
with $r_\Gamma$ and $r_a$ for the typical scale of the bound state vertex and annihilation scale, respectively, which in case of the positronium are
\be
r_\Gamma\simeq\frac{1}{\kappa},\quad \mbox{with}\quad\kappa\simeq \alpha \left(\frac12 m_e\right) = \gamma,
\label{rGamma}
\ee
and
\be
r_a\simeq\frac{1}{m_e}.
\label{rA}
\ee
Note that the case of positronium is very special, since the binding is provided by the massless photon. Accordingly, the effective range of the interaction is provided my $\mu_e \alpha$, which agrees to the binding momentum $\gamma$---clearly, for hadronic molecules where the range of the binding forces is ${\mathcal O}(m_\pi^{-1})$ or shorter, this equality is absent.

Hereinafter, we refer to the hierarchy of Eq.~\eqref{caseA} as case A (thence the subscript $A$ at the wave function) or the ``positronium'' limit. In this limit, the disintegration of the molecule into its constituents effectively takes place at the origin---the point in space where the constituents ``meet'' each other with probability $|\psi_A(0)|^2$. The wave function in Eq.~\eqref{psir} comes as solution of the Schr{\"o}dinger equation with the Coulomb potential, and its Fourier transform takes the form
\be
\psi_A(\vep)=\frac{8\sqrt{\pi\gamma^5}}{(p^2+\gamma^2)^2}.
\label{psipA}
\ee

\begin{figure}[t!]
\centerline{\epsfig{file=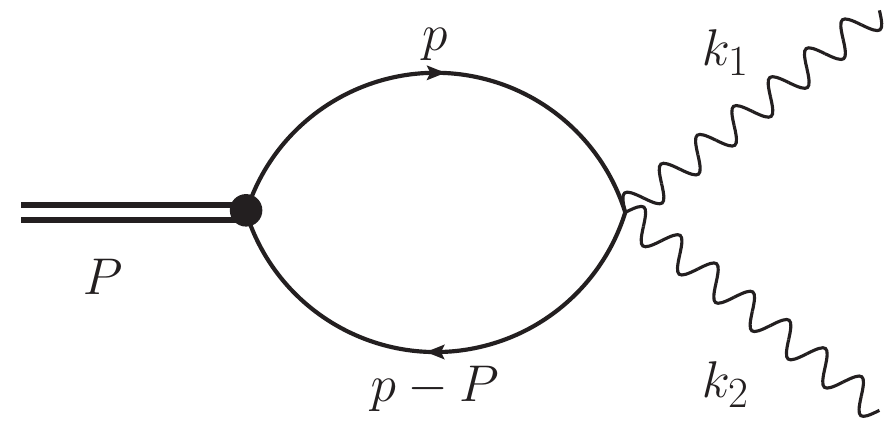,width=0.35\textwidth}}
\caption{Contact diagram that contributes to the amplitude of the two-photon decay of a composite object composed of two charged scalar (or pseudoscalar) particles.}\label{fig:fish}
\end{figure}

As indicated above,
 a different (opposite to case A above) hierarchy of scales is typically relevant for hadronic molecules, namely
\be
r_\Gamma\ll r_a \ ,
\label{caseB}
\ee
which we refer to as case B and accordingly use the corresponding subscript at the wave function. Now the interaction of the constituents responsible for the formation of the molecule is controlled by the mass
of the lightest exchange particle allowed, in
the following denoted as $\beta$. The extreme limit of $\beta\to\infty$ is often referred to as the ``point-like''  or zero-range limit, since then the molecule is formed by a zero-range potential. In this case, the bound state wave function takes the form
\be
\psi_B(r)=\sqrt{\frac{\gamma}{2\pi}}\frac{1}{r}e^{-\gamma r},\quad\gamma=\sqrt{2\mu E_B}, \quad \mu=\frac{m_1 m_2}{m_1+m_2}
\label{psir2}
\ee
in coordinate space, or
\be
\psi_B(\vep)=\frac{\sqrt{8\pi\gamma}}{p^2+\gamma^2},
\label{psipB}
\ee
in momentum space.
More generally, Eq.~\eqref{psir2} describes the long-distance tail of the wave function outside the range where the interaction potential is different from zero. As such this wave function is universal, since it does not depend on the details of the particular potential employed in the corresponding Schr{\"o}dinger equation at short range. It can be obtained as solution of the Schr{\"o}dinger equation in the potential $\delta^{(3)}(\ver)$.
However, in reality it gets modified for distances of the order of the range of forces or lower---a regime that in effective field theories is largely parametrised by local operators (see Ref.~\cite{Nogga:2005fv} for a related discussion).
The fact that the wave function in Eq.~\eqref{psir2} diverges in the limit $r\to 0$ does not cause trouble in
the pertinent calculations, since the positronium formula for the width in Eq.~\eqref{widthA} that involves $\psi(0)$ does not apply, as soon as the interaction is of shorter range than the transition of the constituents to the final state. The described situation is always realized in the point-like limit, such that in case B the two-photon decay width of a molecule of mass $\mS$ is calculated as \cite{Hanhart:2007wa}
\be
\Gamma_{\gamma\gamma}(\beta)
=\Gamma_{\gamma\gamma}\left(1+{\cal O}\left(\frac{mE_B}{\beta^2}\right)\right),\quad
\quad E_B=2m-\mS,
\label{Gexact}
\ee
where
\be
\Gamma_{\gamma\gamma}=\Gamma_{\gamma\gamma}(\beta\to\infty)=\zeta^2\left(\frac{\alpha}{\pi}\right)^2\sqrt{mE_B}\left(\frac{2m}{\mS}\right)
\left[\left(\frac{2m}{\mS}\right)^2\arcsin^2\left(\vphantom{\frac{2m}{\mS}}\frac{\mS}{2m}\right)-1\right]^2,
\label{widthpl}
\ee
and the factor $\zeta\leqslant 1$ accounts for the fact that not all constituents of the molecule (for example, neutral hadrons) may participate in the decay. The expression above is provided for $m_1=m_2=m$ to simplify the formula. Importantly, it was argued on general grounds and demonstrated explicitly in Ref.~\cite{Hanhart:2007wa} that potentially sizable corrections of the order ${\cal O}(m^2/\beta^2)$ to the point-like formula in Eq.~\eqref{widthpl} cannot appear.
Importantly, the width in Eq.~\eqref{widthpl} comes as a result from the straightforward evaluation of the loop amplitude that acquires three contributions: two of them taking the form of the diagrams in Fig.~\ref{fig:posdec} plus a contact diagram depicted in Fig.~\ref{fig:fish} that appears for gauge theories of charged spinless particles interacting with photons as well as for charged non-relativistic particles with spin. The molecular vertex is in an $S$-wave, so it does not
introduce any loop momentum, and the corresponding coupling constant can be evaluated in terms of the binding momentum --- see Eq.~\eqref{vert} below. Thus, all photon emission vertices in the diagrams are electric ones. Then the gauge invariance of the amplitude
leads to delicate cancellations between different terms coming from the above three diagrams and is only in place, if all three contributions to the amplitude are retained.
Those cancellations at the same time remove the divergent pieces from the amplitude, such that the total amplitude is ultraviolet (UV) finite, while its individual (gauge non-invariant!) parts are divergent. Indeed, gauge invariance forces internal loop momenta to get converted into external momenta, this way improving the UV behaviour of the integrated function, to
allow for a transverse tensor structure of the total amplitude.

The formulae in Eqs.~\eqref{Gexact} and \eqref{widthpl} apply to the case of the two-photon decays of the scalar
$f_0(980)$ treated as a $K\bar{K}$ molecule. Indeed, the relevant scales in this case are
\be
r_\Gamma\simeq \frac{1}{\beta}\simeq\frac1{m_\rho},
\quad r_a\simeq\frac{1}{m_K},
\label{rs}
\ee
with $m_\rho\approx 770$~MeV and $m_K\approx 490$~MeV.
It was used in Eq.~\eqref{rs} that the pion exchange between pseudoscalar kaons is forbidden by parity conservation, so the lightest allowed exchange particle is the $\rho$ meson. The hierarchy of the scales in Eq.~\eqref{rs} is obviously at odds with case A in Eq.~\eqref{caseA} and the two-photon decay of a $K\bar{K}$ molecule rather falls into class~B as given by Eq.~\eqref{caseB}. Then Eqs.~\eqref{Gexact} and \eqref{widthpl} give for the width \cite{Hanhart:2007wa}
\be
\Gamma_{\gamma\gamma}^{\rm th}[f_0(980)] = (0.22\pm 0.07)~\mbox{keV},
\label{pw2}
\ee
where $m_S=980$~MeV was used that, together with the kaon mass quoted above, translates into the binding energy $E_B=10$~MeV. It was also taken into account that, in case of
$f_0(980)$, only the charged kaons in the loop couple to the photons in leading order, so $\zeta^2=1/2$. The uncertainty was estimated from the size of the leading neglected correction ${\cal O}(mE_B/\beta^2)$ in Eq.~\eqref{Gexact}, with $\beta=m_\rho$, as given in Eq.~\eqref{rs}.
The result in Eq.~\eqref{pw2} is consistent with the experimental measurement by Belle \cite{Belle:2006wcd},
\be
\Gamma_{\gamma \gamma}^{\rm exp}[f_0(980)]=0.205^{+0.095}_{-0.083}({\rm stat})^{+0.147}_{-0.117}({\rm syst})~{\rm keV}.
\label{newbelle}
\ee

In summary, we stress that the appropriate approach to two-photon decays of composite systems crucially depends on the hierarchy of the scales responsible for the formation of the state and the annihilation of its constituents to a pair of photons.
Since the zero-range potential often provides a valid approximation for the binding interaction of hadronic molecules, a positronium-like treatment, Eq.~\eqref{caseA}, is not justified and the point-like limit of Eq.~\eqref{caseB} should be used instead. In the latter case, the result for the decay width is not sensitive to the details of the interaction at short distances and only depends on the long-range tail of the molecule wave function that is model independent. In case of a $K\bar{K}$ molecule, the loop integral converges in the point-like limit as a result of the gauge invariance of the photon vertices. In particular, no short-range operator needs to be added at leading order, since the kaon loops provide the dominating contribution to the decay amplitude. Therefore, the result in Eq.~\eqref{pw2}
comes as a theoretical prediction for the two-photon decay of the $K\bar{K}$ molecule that depends only on the masses of the decaying state and its constituents.

\section{Radiative decays of $\phi(1020)$}
\label{sec:phi}

\begin{figure*}[t]
\begin{center}
\begin{tabular}{cccc}
\epsfig{file=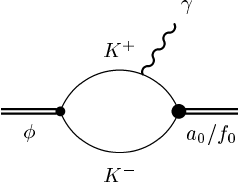,width=3.5cm}&
\raisebox{-6mm}{\epsfig{file=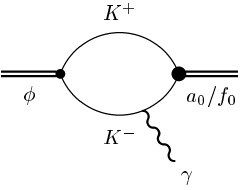,width=3.5cm}}&
\epsfig{file=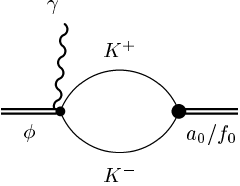,width=3.5cm}&
\epsfig{file=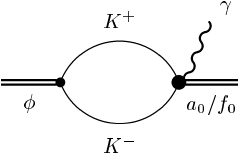,width=3.5cm}\\
(a)&(b)&(c)&(d)
\end{tabular}
\end{center}
\caption{Loop diagrams contributing to the radiative decays $\phi(1020)\to\gamma a_0/f_0(980)$. The diagram (c) guarantees gauge invariance of the total amplitude while the diagram (d) appears only when proceeding beyond the strict point-like limit for the vertex $K\bar{K}\to a_0/f_0$. Adapted from Ref.~\cite{Kalashnikova:2004ta}.}
\label{fig:phirad}
\end{figure*}

The processes $\phi(1020)\to\gamma a_0/f_0(980)$ constitute yet another example of radiative decays involving the scalar mesons $a_0/f_0(980)$. They are studied in Refs.~\cite{Close:1992ay,Achasov:1996ei} and a conclusion is made that, for the scalars treated as $K\bar{K}$ molecules with the radial wave function in Eq.~\eqref{psipA}, the $\phi(1020)$ radiative decay rates are strongly suppressed in comparison with the case of compact scalars. In addition, the authors of Refs.~\cite{Achasov:1996ei,Achasov:2006cq} claim that the integrals in the corresponding loop amplitudes in Fig.~\ref{fig:phirad} are sensitive to very large (of the order of GeV's) loop momenta and, therefore, $a_0/f_0(980)$ need to be interpreted as compact tetraquark states rather than extended hadronic molecules. These findings and conclusions are scrutinised and criticised in Refs.~\cite{Kalashnikova:2004ta,Kalashnikova:2007dn}, so
below we briefly mention the arguments put forward in the latter works
that are in line with the discussion above.

It was shown in the previous section that the scalars $a_0/f_0(980)$ treated as $K\bar{K}$ molecules comply with the hierarchy of the relevant scale in Eq.~\eqref{caseB} rather than Eq.~\eqref{caseA}, so the choice of the wave function in the form of $\psi_A(\vep)$ in Eq.~\eqref{psipA} is not justified. As a next step we derive the molecule disintegration vertex and notice that, in the vicinity of a bound state, the nonrelativistic $t$-matrix takes the form
\be
t(\vep,\vep',E)=\frac{f(\vep)f(\vep')}{E+E_B-i0},
\label{t}
\ee
where the nonrelativistic vertex $f(\vep)=\hat{v}\psi_B(\vep)$ (here $\psi_B(\vep)$ is the wave function in Eq.~\eqref{psipB} and $\hat{v}$ is the nonrelativistic interaction operator) can be found directly from the Schr{\" o}dinger equation in momentum space,
\be
\frac{\vep^2}{2\mu}\psi_B(\vep)-\hat{v}\psi_B(\vep)=-E_B\psi_B(\vep),
\label{eq}
\ee
with $E_B=\gamma^2/(2\mu)$, that gives a constant nonrelativistic vertex,
\be
f(\vep)=\frac1{2\mu}(\vep^2+\gamma^2)\psi_B(\vep)=\frac{\sqrt{8\pi\gamma}}{2\mu},
\label{vert}
\ee
in agreement with the point-like limit introduced in the previous section and the famous Weinberg formula for the molecule coupling to its constituents \cite{Weinberg:1965zz,Guo:2017jvc}. Explicit calculations performed in
Ref.~\cite{Kalashnikova:2004ta} demonstrate that, for a realistic set of parameters, there is no suppression of the decay rates $\phi(1020)\to\gamma a_0/f_0$ for the scalar mesons regarded as molecules, and
the deviations from the point-like limit caused by a residual momentum dependence of the molecular vertex have only very little effect on the result. It needs to be noted that, similarly to the case of the two-photon decay of a molecule addressed in the previous section, the photon is emitted from electric vertices and the individual loop diagrams in Fig.~\ref{fig:phirad}(a)-(c) lead to divergent integrals but the full amplitude that sums these contributions is finite. Moreover, contrary to the claim in Refs.~\cite{Achasov:1987ts,Achasov:2006cq}, after a proper treatment of the individual contributions to the amplitude, the resulting loop integral converges at nonrelativistic momenta---again a
direct consequence of gauge invariance.
We therefore conclude that the data for the scalars $a_0/f_0(980)$ in the final state of the $\phi(1020)$ radiative decay are consistent with these
enigmatic states being pure $\bar KK$
molecular states.

\section{Radiative decays of $\dsone(2460)$}
\label{sec:Ds1}

In this section, we review the radiative decays of the $\dsone(2460)$. We start from the two-body decay $\dsone(2460)\to \gamma \dszero(2317)$ assuming dominating molecular components in the wave functions of both $D_{sJ}$ mesons
involved~\cite{Barnes:2003dj,Kolomeitsev:2003ac,Chen:2004dy,Guo:2006fu,Guo:2006rp,Gamermann:2006nm,Lutz:2007sk,Faessler:2007gv}.
The effective Lagrangian describing this decay takes the form
\be
{\cal L}_{\rm eff}=\kappa\varepsilon^{\mu\nu\alpha\beta}F_{\mu\nu}v_{\beta}D_{s1\alpha}D^{*\dagger}_{s0}+\textrm{h.c.},
\label{LagDs1Gsogamma}
\ee
where $v$ denotes the 4-velocity of the $\dsone$ meson and $\kappa$ represents a dimensionless effective coupling constant. Notice that an alternative
identification of $v$ as the 4-velocity of the decaying $\dszero$ meson would only affect subleading terms in the heavy quark mass expansion that can be neglected. Then the decay width is calculated as
\be
\Gamma[\dsone\to\gamma \dszero]=\frac{\kappa^2\omega^3}{3\pi m_{\dsone}^2},
\label{WidthDs1Gsogamma}
\ee
where $\omega\approx 139$~MeV is the energy of the photon.

\begin{figure}[t!]
\centering
\epsfig{file=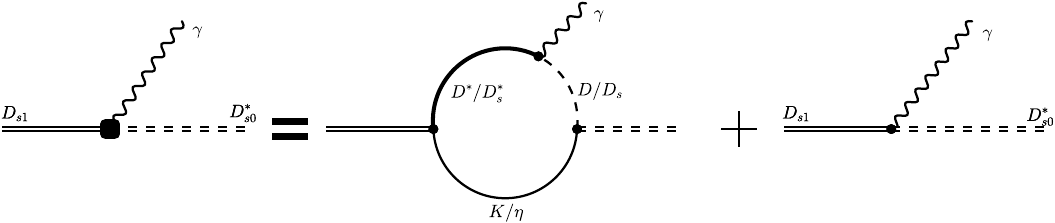, width=0.8\textwidth}
\caption{The loop and contact diagrams contributing to the radiative decay $\dsone(2460)\to \gamma \dszero(2317)$. Adapted from \cite{Fu:2025lfo}.}\label{fig:Ds1Ds0gamma}
\end{figure}

\begin{figure}
\centering
\includegraphics[width=0.7\linewidth]{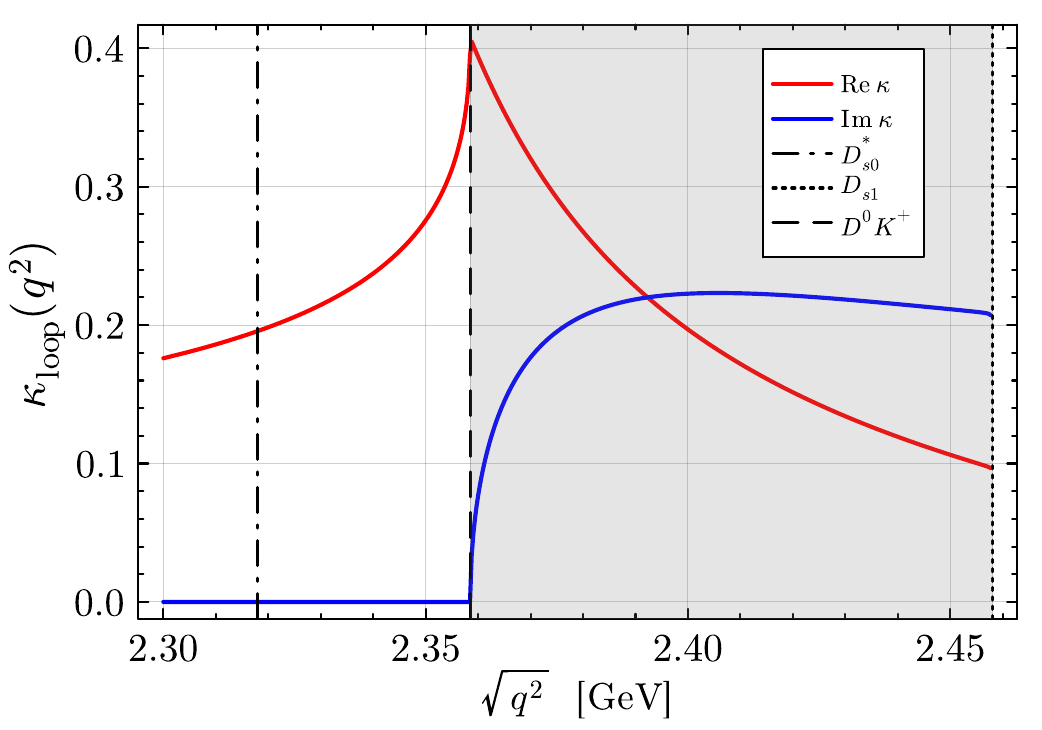}
\caption{The momentum dependence of the effective loop coupling $\kappa_{\rm{loop}}(q^2)$, as an example, evaluated for the masses of $D^{(*)+}$ and $K^0$.
The vertical dashed-dotted line shows the position
of $q^2=m_{\dszero}^2$ relevant for the two-body decay $\dsone(2460)\to\gamma\dszero(2317)$. The gray band shows the
range of the phase space integration in the three-body decay $\dsone(2460)\to\gamma D^0K^+$,
$(m_{D^{0}}+m_{K^{+}})^2\leqslant q^2\leqslant m_{\dsone}^2$.
Adapted from Ref.~\cite{Fu:2025lfo}.}
\label{fig:kappaq2}
\end{figure}

Formally, the decay $\dsone(2460)\to \gamma\dszero(2317)$ proceeds through the diagrams shown in Fig.~\ref{fig:Ds1Ds0gamma}. We study the loop amplitude first and notice that (i) the molecular vertices $\dsone\to D^*K$ and $\dszero\to DK$ are $S$-wave ones, so they do not bring powers of the loop momentum and (ii) the coupling constants can be evaluated either from the corresponding binding energies using the well-known Weinberg formula~\cite{Weinberg:1965zz} or from the residues of the scattering amplitudes in the unitarised chiral perturbation theory (UChPT) that generates the $\dszero$ and $\dsone$ mesons dynamically. In addition, the magnetic photon emission vertex $D^*\to\gamma D$ is also well understood and its parameters can be extracted directly from the experimental data on the corresponding radiative decay. Then,
as discussed in Ref.~\cite{Fu:2025lfo}, the momentum-dependent effective loop coupling $\kappaoff$ coming from the loop diagram in Fig.~\ref{fig:Ds1Ds0gamma} can be evaluated straightforwardly since the corresponding loop amplitude involves only convergent integrals.
It is instructive to note that, in this case, gauge invariance (and UV-finiteness) of the loop amplitude is guaranteed by the anomalous coupling in the effective Lagrangian \eqref{LagDs1Gsogamma} and the magnetic photon emission vertex $D^*\to\gamma D$ that is by construction transverse with respect to the photon momentum.
In particular, here the subtle cancellations discussed above to not take place.
For a realistic set of parameters, $\kappaoff$ evaluated on the mass shell of the $\dszero$ reads \cite{Fu:2025lfo},
\be
\kappaon\approx 0.19.
\label{kappaloop}
\ee
The momentum dependence of $\kappaoff$ is depicted in Fig.~\ref{fig:kappaq2}, with the result in Eq.~\eqref{kappaloop} shown by the vertical dashed-dotted line.

\begin{figure*}[t]
\centering
\begin{tabular}{ccc}
\includegraphics[width=0.3\textwidth]{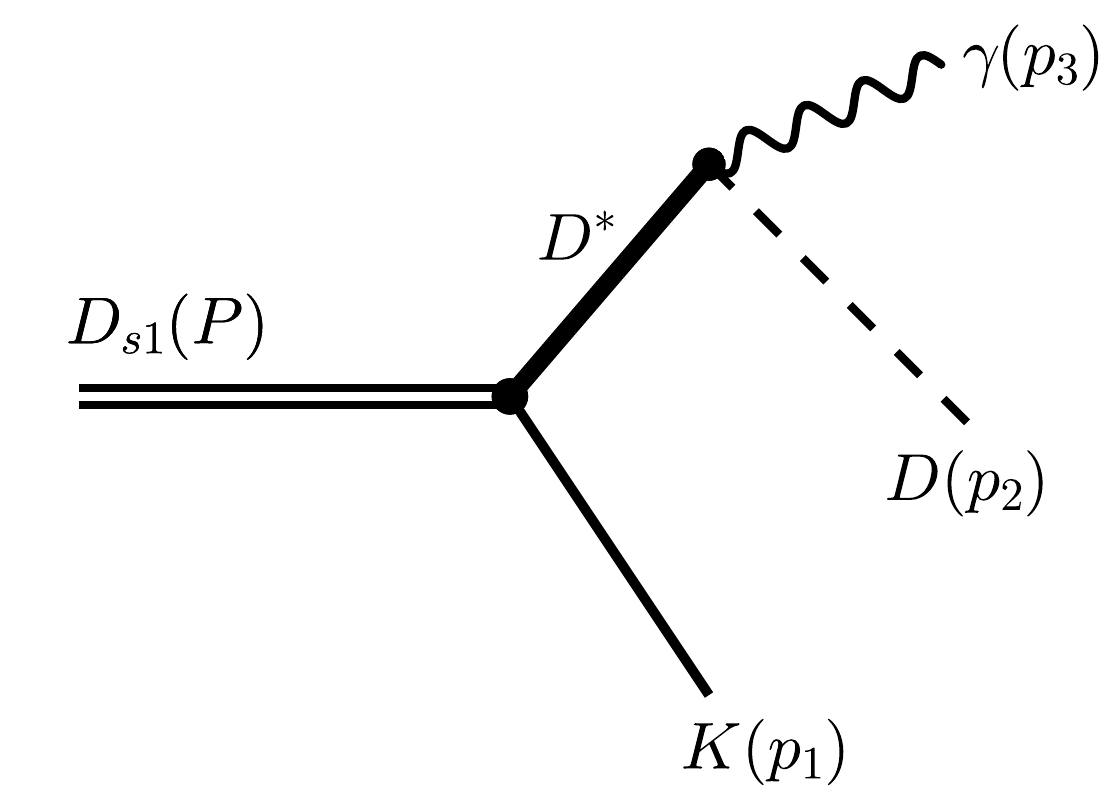} & \includegraphics[width=0.3\textwidth]{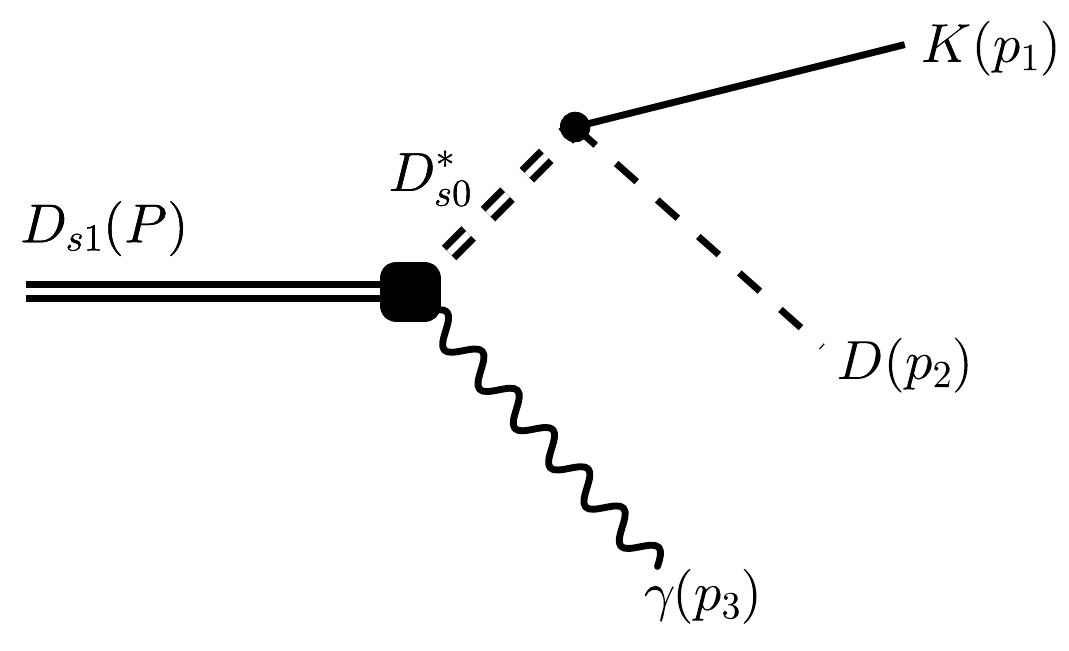} &
\includegraphics[width=0.3\textwidth]{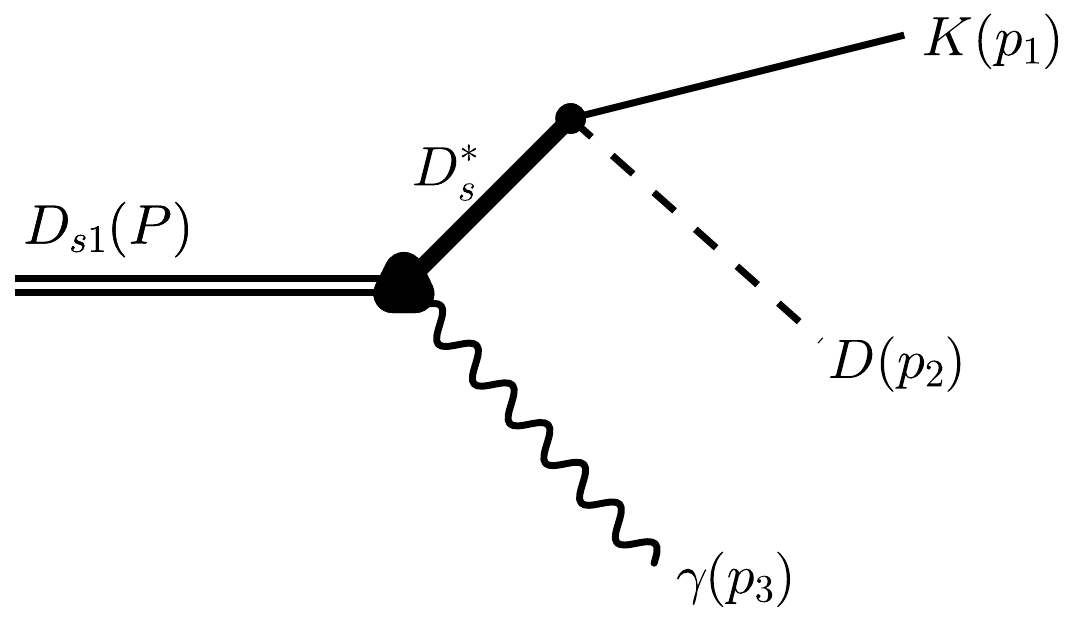}
\\
(a) & (b) & (c)
\end{tabular}
\caption{Various contributions to the decay amplitude $\dsone\to \gamma DK$. The structure of the vertex $\dsone\to \gamma \dszero$ in diagram (b) is shown in Fig.~\ref{fig:Ds1Ds0gamma}. Details concerning the vertex $\dsone\to \gamma D_{s}^*$ in diagram (c) can be found in Refs.~\cite{Cleven:2013rkf,Cleven:2014oka,Fu:2021wde}. Adapted from Ref.~\cite{Fu:2025lfo}; see this reference also for a detailed investigation of the corresponding amplitude}.
\label{fig:Ds1DKgamma}
\end{figure*}

As argued in Ref.~\cite{Fu:2021wde} and contrary to the hadronic decays of the $\dsone(2460)$, there is no enhancement of the loop diagrams compared with the contact term in its radiative decay. Then the full coupling $\kappa$ in Eq.~\eqref{WidthDs1Gsogamma} acquires two contributions,
\be
\kappa=\kappaon+\kappa_{\rm cont},
\label{kappadef}
\ee
where the term $\kappa_{\rm cont}$ sets the strength of the contact diagram in Fig.~\ref{fig:Ds1Ds0gamma}. The value of $\kappa_{\rm cont}$ is currently unknown and was estimated in Ref.~\cite{Fu:2025lfo} in two ways. On the one hand,
Eq.~\eqref{WidthDs1Gsogamma} was applied to a quark-model result for the decay width of a generic $c\bar{s}$ state,
$\Gamma[1^+(c\bar{s})\to 0^+(c\bar{s})+\gamma]\approx 2.74$~keV~\cite{Bardeen:2003kt}, to find
\be
|\kappa_{\rm cont}|\simeq 0.24.
\label{kappacont1}
\ee
On the other hand, the exciting experimental data on the radiative decay $\dsone(2460)\to \gamma \dszero(2317)$~\cite{ParticleDataGroup:2024cfk} and theoretical estimates for various $\dsone(2460)$ partial decay widths were employed to get \cite{Fu:2025lfo}
\be
\kappa_{\rm cont}\simeq 0.2.
\label{kappacont2}
\ee
Remarkably, the numerical estimates in Eqs.~\eqref{kappacont1} and \eqref{kappacont2} comply well with both a natural expectation $\kappa_{\rm cont}\simeq\Lambda_{\rm QCD}/m_c\simeq 0.2$, coming from the fact that the considered decay involves a heavy quark spin flip, and the power counting argued in Ref.~\cite{Cleven:2014oka}, thus yielding
\be
|\kappa_{\rm cont}|\simeq|\kappaon|.
\ee
Then indeed, as anticipated in Ref.~\cite{Fu:2021wde}, different models for the structure of the $\dsone$ meson may lead to similar results for its radiative decay width.
However, the result in Eq.~\eqref{kappacont1} is strongly model dependent while, given the very poor knowledge of the experimental branching $\mbox{Br}(\dsone(2460)\to \gamma \dszero(2317))$, the estimate in Eq.~\eqref{kappacont2} comes with a large uncertainty, around 100\%. Thus, more reliable and accurate methods to evaluate $\kappa_{\rm cont}$ need to be invoked. In particular, it is argued in Ref.~\cite{Fu:2025lfo} that the value of the contact contribution $\kappa_{\rm cont}$ could be extracted from the ratio
\be
{\mathcal R}_{\dsone}=\frac{\mbox{Br}(\dsone(2460)\to \gamma D^{*}_{s0}(2317))}{\mbox{Br}(\dsone(2460)\to \gamma D^0K^+)}
\label{ratioR}
\ee
that demonstrates a lively dependence on this parameter. The amplitude for the transition $\dsone(2460)\to \gamma \dszero(2317)$ enters the amplitude for the three-body decay $\dsone(2460)\to \gamma D^0K^+$ as a substructure (see Fig.~\ref{fig:Ds1DKgamma}(b)) and in this way gives it a dependence on $\kappa_{\rm cont}$.
Experimental determination of the ratio in Eq.~\eqref{ratioR} or at least imposing a constraint on its value, which is easier than measuring the involved partial decay widths separately, may already be sufficient to arrive at a fair estimate for $\kappa_{\rm cont}$. Furthermore, as found in Ref.~\cite{Fu:2025lfo}, the width of the three-body radiative decay $\dsone(2460)\to \gamma D^0K^+$ is sensitive to the loop coupling $\kappaoff$ evaluated in a different kinematical regime compared with the two-body decay $\dsone(2460)\to \gamma D^{*}_{s0}(2317)$ --- see the gray shaded band in Fig.~\ref{fig:kappaq2}. It is, therefore, further argued in Ref.~\cite{Fu:2025lfo} that experimental studies of this three-body radiative decay may shed light on the internal structure of the $D_{sJ}$ mesons involved.

We, therefore, conclude this section noting that the radiative decays of the $\dsone(2460)$ are sensitive to both the long-range molecular and short-range compact components of its wave function which can potentially be quantified if sufficiently accurate experimental data become available.

\section{Radiative decays of the $\X$}
\label{sec:X}

In this section, we discuss the radiative decays $\X\to\gamma \psi(2S)$ and $\X\to \gamma J/\psi$ that, in recent years, attracted a lot of attention of both experimentalists and theorists. An interest to these decays was recently catalysed by the improved analysis performed by the LHCb Collaboration. In particular, it provided the most precise measurement so far
of the ratio
\be
{\cal R}_X=\frac{\mbox{Br}(\X\to\gamma \psi(2S))}{\mbox{Br}(\X\to \gamma J/\psi)}
\label{R}
\ee
that reads \cite{LHCb:2024tpv}
\be
{\cal R}_X^{\rm LHCb}=1.67\pm 0.21\pm 0.12\pm 0.04
\label{RLHCb}
\ee
and supersedes the previous, 10-year-old result belonging to the same collaboration \cite{LHCb:2014jvf}.
The ratio in Eq.~\eqref{R} has been under an intensive discussion in the literature starting from its first experimental measurement by the BaBar Collaboration in 2008 \cite{BaBar:2008flx}. Since then, several attempts were undertaken to improve on this value (see Table~\ref{tab:rad} for a collection of the measurements belonging to various experimental groups), and the recent measurement by LHCb in Eq.~\eqref{RLHCb} indeed considerably advances the experimental situation.

From the theory side, shortly after the $\X$ discovery by the Belle collaboration in 2003~\cite{Belle:2003nnu}, the ratio of its radiative decays in Eq.~\eqref{R} was argued to provide a decisive selection rule to discriminate between different assignments for the nature of the $X$ \cite{Swanson:2004pp}. In particular, for this state interpreted as a pure $D\bar{D}^*$ molecule, the cited work reported the expectation ${\cal R}_X\ll 1$ at odds with most of the experimental results in Table~\ref{tab:rad}. Shortly after the updated LHCb analysis in Ref.~\cite{LHCb:2024tpv} was announced, a tiny ratio ${\cal R}_X$
for a molecular $\X$ was argued for in Ref.~\cite{Grinstein:2024rcu}, based on the estimates of the overlaps of various wave functions with the molecular wave function taken to be the asymptotic long-distance form as in Eq.~\eqref{psir2}.
This leads to the conclusion that the recent LHCb measurement precludes a molecular nature of the $\X$.
After some general remarks concerning the $\X$, we demonstrate below that these claims
are at odds with the general discussion presented in the previous sections.

\begin{table}[t!]
\caption{Experimental measurements of the ratio ${\cal R}_X$ in Eq.~\eqref{R} for the $\X$ radiative decays widths.}
\label{tab:rad}
\begin{tabularx}{\textwidth}{CCCCC}
\toprule
BaBar \cite{BaBar:2008flx} & Belle \cite{Belle:2011wdj} & LHCb (old) \cite{LHCb:2014jvf} & BESIII \cite{BESIII:2020nbj} & LHCb (new) \cite{LHCb:2024tpv}\\
\midrule
$3.4\pm 1.4$ & $<2.1$ & $2.46\pm 0.64\pm 0.29$ & $<0.59$ & $1.67\pm 0.21\pm 0.12\pm 0.04$\\
\bottomrule
\end{tabularx}
\end{table}

Given the unambiguously established quantum numbers $1^{++}$ of the $\X$ \cite{Aaij:2013zoa,Aaij:2015eva} and a proximity of its mass to the nominal neutral $D\bar{D}^*$ threshold \cite{ParticleDataGroup:2024cfk}, assuming a considerable molecular $D\bar{D}^*$ component in the $X$ wave function looks quite plausible.\footnote{Hereinafter, a proper $C$-even combination of the $D\bar{D}^*$ and $\bar{D}D^*$ components is understood.} Then, based on the experience gained in the previous sections, a relevant question that arises is whether the hierarchy of the intrinsic scales in the $\X$
is more consistent with the positronium case in Eq.~\eqref{caseA} or the point-like limit in Eq.~\eqref{caseB}. It should be noted that, unlike the system of two kaons studied above, the pion exchange is not forbidden
between the $D$ and $\bar{D}^*$ mesons. Meanwhile, the $DD\pi$ vertex is still not allowed by $P$-parity conservation, so the pion-driven interaction between the $D$ and $\bar{D}^*$ proceeds in the $u$-channel, and the corresponding potential reads
\be
V_\pi(\vep,\vep')
=3\left(\frac{g_c}{2f_\pi}
\right)^2
\frac{(\veep\cdot \veq)(\veq\cdot {\veep^{\prime *}})}{u-m_\pi^2},
\label{VOPE}
\ee
where $\vep^{(\prime)}$ and $\veep^{(\prime)}$ are
the momenta and polarisation vectors of the $D^*$ mesons in the initial and final state, respectively, $\veq=\vep'-\vep$ is the pion momentum and the factor 3, stems from the eigenvalue of the isospin operator ${\bm\tau}\cdot{\bm\tau}$ for an isoscalar $D\bar{D}^*$ system (we work in the strict isospin limit --- for a discussion on the role of isospin violation and how to find a possibly existing isovector partner state see Ref.~\cite{Ji:2025hjw}). Furthermore,
$$
u-m_\pi^2=q^2-m_\pi^2=(q_0^2-\veq^2)-m_\pi^2=-\veq^2-\left(m_\pi^2-q_0^2\right)
\approx-(\veq^2+\mu_\pi^2),
$$
where the recoil terms for the $D^{(*)}$ mesons were neglected by setting
$q_0= E_{\bar{D}^*}-E_D\approx m_{D^*}-m_D$\footnote{In this way the three-body cut, discussed in some detail in Ref.~\cite{Baru:2011rs}, is removed from the amplitude, however, this is not crucial for the discussion here.} and an effective pion mass parameter,
\be
\mu_\pi^2=m_\pi^2-(m_{D^*}-m_D)^2,
\label{mupi}
\ee
has appeared naturally~\cite{Tornqvist:1993ng,Swanson:2003tb,Suzuki:2005ha}. Numerically $|\mu_\pi|\approx 42$~MeV and one may be tempted to deduce from this estimate that, in the $D\bar{D}^*$ system at hand, the inverse range of the force $\beta$ is given by the parameter $|\mu_\pi|$ and as such it is small compared with the other mass scales present in the system. Then, na{\"i}vely, the positronium-like hierarchy of the scales in Eq.~\eqref{caseA} should be relevant for the $\X$. However, the latter conclusion is not justified. To demonstrate this, consider the central part of the pion exchange potential \eqref{VOPE},
\be
V_\pi^{\rm cent}(\veq)=\left(\frac{g_c}{2f_\pi}
\right)^2
\frac{\veq^2}{u-m_\pi^2}= \left(\frac{g_c}{2f_\pi}
\right)^2\left(-1+\frac{\mu_\pi^2}{\veq^2+\mu_\pi^2}\right),
\label{Vsl}
\ee
where the substitution $\epsilon_i\epsilon^{\prime*}_j\to\frac13\delta_{ij}$, relevant for the central potential, was used. Since a constant potential in momentum space (coming from the term $-1$ in parentheses) turns to a zero-radius potential $\delta^{(3)}(\ver)$ in coordinate space, Eq.~\eqref{Vsl}
corresponds to an explicit separation of the short- and long-ranged contributions, and the second term in parentheses vanishes in the limit $|\veq|\to\infty$. Therefore, as stressed in Ref.~\cite{Baru:2015nea}, the OPE potential in the $D\bar{D}^*$ system is well defined in the sense of an effective field theory only in connection with a contact operator that corresponds to the range parameter $\beta\to\infty$.
\begin{figure*}
\begin{center}
\begin{tabular}{ccc}
\raisebox{13mm}{(a)}~\epsfig{file=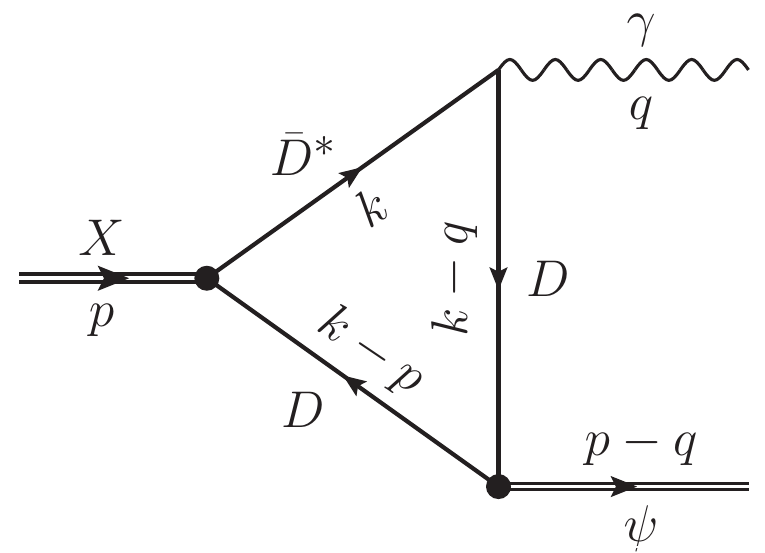,width=0.25\textwidth}&
\raisebox{13mm}{(b)} ~\epsfig{file=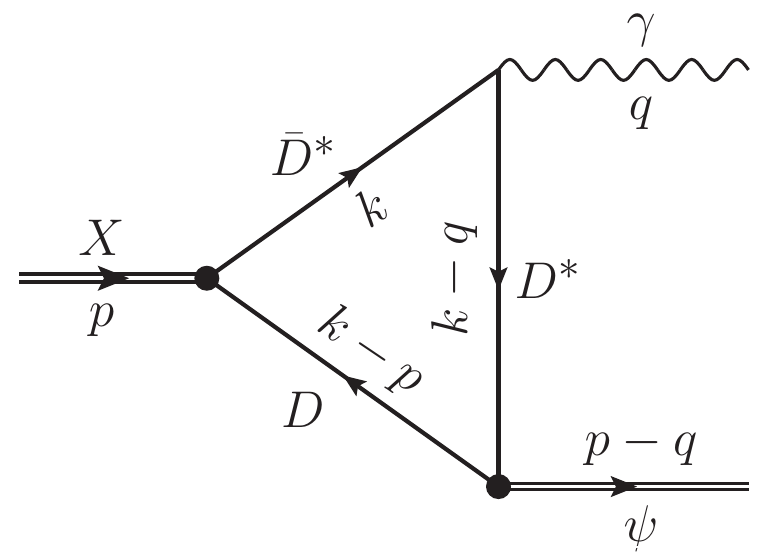,width=0.25\textwidth}&
\raisebox{13mm}{(c)} ~\epsfig{file=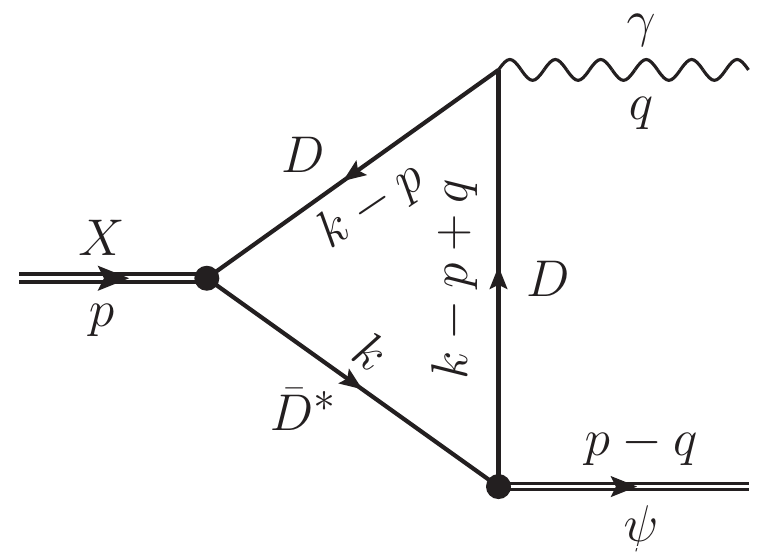,width=0.25\textwidth}\\
\raisebox{13mm}{(d)} ~\epsfig{file=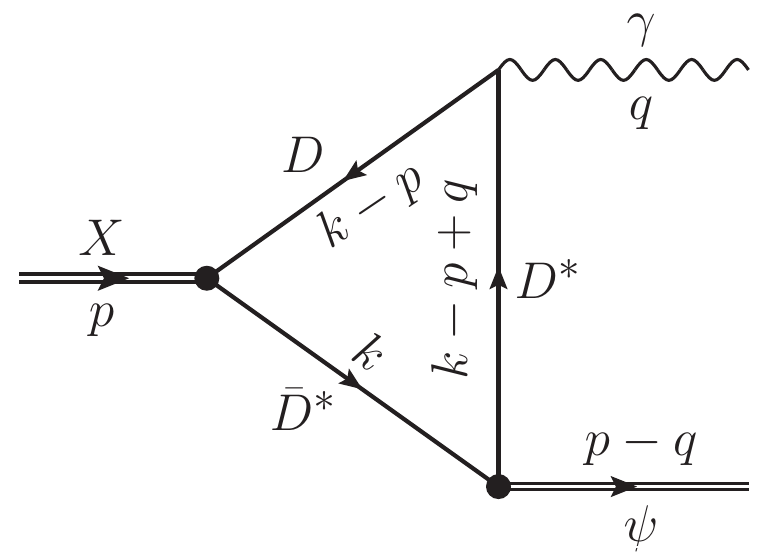,width=0.25\textwidth}&
\raisebox{13mm}{(e)} ~\raisebox{2.5mm}{\epsfig{file=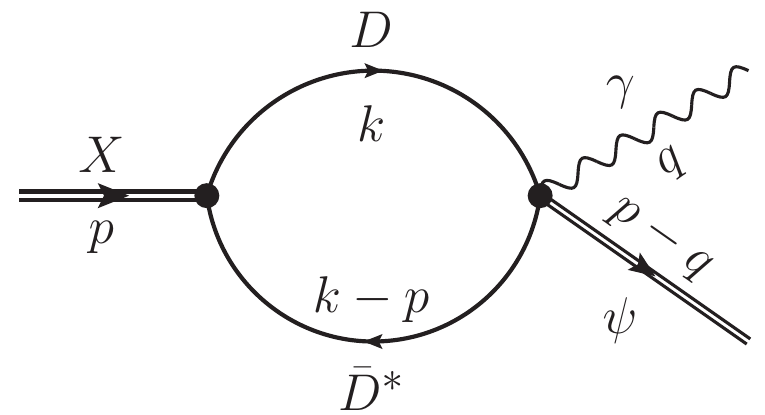,width=0.25\textwidth}}&
\raisebox{13mm}{(f)} ~\raisebox{5.5mm}{\epsfig{file=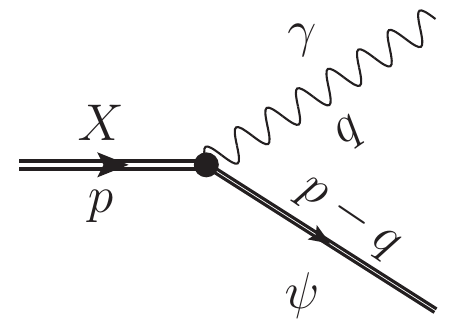,width=0.15\textwidth}}
\end{tabular}
\end{center}
\caption{The diagrams contributing to the amplitude of the decay $\X\to\gamma\psi$ with $\psi=J/\psi$, $\psi(2S)$. Adapted from Ref.~\cite{Guo:2014taa}.}
\label{fig:Xrad}
\end{figure*}
The latter may acquire additional contributions from various heavy-meson exchanges that are allowed between the $D$ and $\bar{D}^*$ mesons.
Therefore, the hierarchy of the scales relevant to the radiative decays of the $\X$ is in fact provided by case B in Eq.~\eqref{caseB}. Following the discussion in Sec.~\ref{sec:phi} above and substituting the momentum-space wave function in Eq.~\eqref{psipB} into the vertex in Eq.~\eqref{vert},
one concludes that the decay vertex $\X\to D\bar{D}^*$ can be taken as a constant that drops out from the ratio in Eq.~\eqref{R}. Thus, no overlap of the deuteron-like wave function in Eq.~\eqref{psir2} with anything else can drive the ratio ${\cal R}_X$ in Eq.~\eqref{R} as a matter of principle, contrary to the claim in Ref.~\cite{Grinstein:2024rcu}. Instead, the loop amplitudes depicted in Fig.~\ref{fig:Xrad}(a)-(e) need to be evaluated. The result of this calculation reported in Ref.~\cite{Guo:2014taa} (see also an update contained in Ref.~\cite{Shi:2023mer} and a similar calculation in Ref.~\cite{Molnar:2016dbo})
\begin{itemize}
\item[(i)] shows that the mechanism of the radiative decays of the $\X$, assumed to be a $D\bar{D}^*$ molecule, does not imply light-quark pair annihilation, as considered in Ref.~\cite{Grinstein:2024rcu}, since the photon is emitted electrically or magnetically from a $D^{(*)}$ meson in the loop;
\item[(ii)] implies that, even for a purely $D\bar{D}^*$ molecule assignment for the $\X$, the ratio in Eq.~\eqref{R} can be of the order of unity under a not restrictive assumption for the couplings of the vector charmonia $J/\psi$ and $\psi(2S)$ to the $D^{(*)}$ mesons being numerically close to each other;\footnote{Since both $J/\psi$ and $\psi(2S)$ lie below the open-charm threshold, these couplings cannot be evaluated from the corresponding decay widths. Meanwhile, since the $\psi(2S)$ has a larger mass and therefore lies closer to the relevant open-charm thresholds, the corresponding couplings are expected to somewhat exceed those inherent in the $J/\psi$ that lies lower in the spectrum of charmonium and, therefore, further from the open-charm thresholds.}
\item[(iii)] demonstrates a sizeable dependence of the ratio in Eq.~\eqref{R}
on the renormalisation scale $\mu_{\rm ren}$ introduced by the regularisation procedure applied to the divergent loop amplitudes. In particular,
\be
\delta R_{\rm mol}(\mu_{\rm ren})\simeq R_{\rm mol}(\mu_{\rm ren}),
\label{Rmol}
\ee
for $\mu_{\rm ren}$ varied in the range from $m_X/2$ to $2m_X$, with $m_X$ for the nominal $X$ mass.
\end{itemize}
The latter conclusion above is naturally interpreted then as a signal of sensitivity of the ratio ${\cal R}_X$ to the short-range physics encoded in the additional contact diagram depicted in Fig.~\ref{fig:Xrad}(f). The corresponding short-range interaction needs to absorb the dependence of the decay amplitude on $\mu_{\rm ren}$ to ensure that the total result is consistent with the renormalisation group equation $\partial {\cal R}_X/\partial\mu_{\rm red}=0$. In particular, Eq.~\eqref{Rmol} implies that the contribution from the compact component given by the diagram in Fig.~\ref{fig:Xrad}(f) should be at least of the same size as the ``molecular'' contribution coming from the diagrams in Fig.~\ref{fig:Xrad}(a)-(e). However, no further quantitative conclusion on the nature of the $\X$ can be deduced from this calculation, since a scheme-independent separation of the contributions to the ratio in Eq.~\eqref{R} coming from the compact and molecular components is not possible as a matter of principle. Thus one is forced to conclude that the radiative decays of the $\X$ are sensitive to the short-range component of the $X$ wave function and as such are out of control in the effective field theory framework for the $X$. In other words, the ratio ${\cal R}_X$ in Eq.~\eqref{R} is not decisive in establishing the nature of the $\X$.\footnote{Assuming the radiative decays of the $\X$ to proceed through the short-range component of its wave function one can naturally arrive at the experimental ratio in Eq.~\eqref{R} --- see, for example, Refs.~\cite{Guo:2014taa,Brambilla:2024imu}.} One the other hand, the radiative decays $Y(4260)\to \gamma \X$ that involve the $\X$ in the final state are sensitive to its molecular component~\cite{Guo:2013zbw}.

To conclude this section, it is instructive to compare the case of the $X$ radiative decays studied here to the decays addressed in the previous sections. On the one hand, unlike the $D_{s1}\to\gamma D_{s0}^*$ decay in Sec.~\ref{sec:Ds1}, only one molecular state is involved in the process $X\to\gamma\psi$.
Moreover, the charmonium $\psi$ is not selective in
its coupling to the various $D^{(*)}\bar{D}^{(*)}$ pairs, which appears to be in $P$-wave and therefore brings an extra power of the loop momentum. As a result, photon emission vertices of both electric and magnetic types contribute to the total amplitude. In the meantime, to preserve parity, the coupling of the $X$ to $\gamma\psi$ has to be anomalous and involves the field tensor for the electromagnetic field,
\be
{\cal L}_{\rm eff}=g_{X\psi\gamma}\epsilon_{\mu\nu\sigma\lambda}X^\mu\psi^\nu F^{\sigma\lambda}.
\ee
It automatically guarantees gauge invariance of the amplitude and, therefore, the latter does not call for the delicate cancellations between the different parts of the amplitude  to form a transverse structure as discussed in Sec.~\ref{sec:pos}. As a result, the amplitude of the decay $X\to\gamma\psi$ diverges
while staying consistent with gauge invariance.
This marks the crucial difference
of the radiative $X$
decays
with the cases of the radiative decays that were addressed in Secs.~\ref{sec:pos} and \ref{sec:phi}.

\section{Conclusions}
\label{sec:concl}

In this paper, we reviewed various radiative decays involving hadronic molecules and stressed that information on such decays can provide valuable insight into their nature. However, one has to bear in mind that
\begin{itemize}
\item[(i)] the hierarchy of the scales relevant to the studied hadron plays a crucial role --- sticking to a wrong hierarchy will result in a misinterpretation of the information contained in the experimental data;
\item[(ii)] different radiative decays may be sensitive to different components of the wave function of the studied hadron, so probing its long-range molecular component requires a careful choice of the observables to employ.
\end{itemize}

We provided several examples of the radiative decays that involve various strong candidates to hadronic molecules illustrating the general pattern just outlined. In particular, we argue that a typical hierarchy of the scales inherent in hadronic molecules precludes employing the
positronium-like formula based on $\psi(0)$ for their two-photon decays. Instead, the molecule disintegration vertex to its constituents should be taken as a constant obtained either employing the Weinberg formula or, more generally, derived from the residue of the corresponding scattering amplitude at the pole. After that, the loop diagrams that arise for a considered radiative decay need to be evaluated employing the standard field theory technique. Then one of the following three situations can take place:
\begin{itemize}
\item The decay amplitude contains convergent loop integrals and the power counting shows no need in a contact diagram.
The width of such radiative decay can be evaluated without unknown parameters and then, as a prediction, confronted with the existing experimental data. This way the data allow one to probe the molecular nature of the hadronic state under study. The given situation is well exemplified with the radiative decays $S\to\gamma\gamma$ and $\phi(1020)\to\gamma S$ (with $S=a_0/f_0(980)$ treated as $K\bar{K}$ molecules) discussed in Secs.~\ref{sec:pos} and \ref{sec:phi}.
\item The decay amplitude contains convergent loop integrals but the power counting demonstrates no enhancement of the loop diagrams over a contact one. Then a contact term needs to be added to the effective Lagrangian with an unknown coefficient. In this case, experimental data need to be employed first to fix the strength of the contact interaction. The radiative decay
$\dsone(2460)\to \gamma D^{*}_{s0}(2317)$, with both $D_{sJ}$ mesons treated as $D^{(*)}K$ molecules, addressed in Sec.~\ref{sec:Ds1} provides an example of this pattern. It is expected that, once the strength of the contact interaction is fixed from the experiment, further predictions become possible for various reactions involving the vertex $\dsone\dszero\gamma$ as a building block or related to it via heavy quark symmetry.
\item The decay amplitude contains divergent loop integrals and, therefore, requires a contact diagram already at leading order. The strength of the corresponding contact term can be estimated from its scheme dependence and, if the latter is strong, one is forced to conclude that the studied radiative decay is sensitive to the short-range component of the wave function of the decaying state rather than its long-range molecular one. In this case, experimental studies of this radiative decay do not allow one to probe the molecular nature of the state in question. Such situation is well exemplified by the radiative decays $X(3872)\to\gamma\psi$ (with $\psi=J/\psi$, $\psi(2S)$) addressed in Sec.~\ref{sec:X}.
\end{itemize}

We conclude, therefore, that (i) identification of the hierarchy of the scales inherent to the studied molecule candidate, (ii) establishing the power counting relevant to various contributions to its decay amplitude, and (iii) verification of the convergence and scheme dependence of the loop integrals involved are the
necessary prerequisites that allow one to unambiguously judge whether or not the studied radiative decays can probe the nature of the given hadronic state.

\funding{This work is supported in part by the National Key R\&D Program of China under Grant No. 2023YFA1606703;
by the Chinese Academy of Sciences (CAS) under Grant No.~YSBR-101; and by the National Natural Science Foundation of China (NSFC) under Grants No. 12125507, No. 12361141819, and No. 12447101.
Work of A.N. was supported by Deutsche Forschungsgemeinschaft (Project No. 525056915). A.N. and C.H. also acknowledge the support from the CAS President's International Fellowship Initiative (PIFI) (Grants No.~2024PVA0004\_Y1 and No.~2025PD0087).}

\conflictsofinterest{The authors declare no conflicts of interest.}

\dataavailability{No new data were used or produced in this research.}


\end{document}